\documentclass[12pt]{iopart}
\usepackage{iopams}
\usepackage[xdvi]{graphicx}

\newcommand{\bra}{\left\langle}
\newcommand{\ket}{\right\rangle}
\newcommand{\pder}[2]{\frac{\partial #1}{\partial  #2}}
\newcommand{\pdert}[3]{\frac{\partial^2 #1}{ {\partial  #2}{\partial #3}}}
\newcommand{\der}[2]{\frac{d #1}{d  #2}}

\newcommand{\lr}[1]{\left( #1 \right)}
\newcommand{\nm}{\nonumber\\}

\newcommand{\bv}[1]{{\boldsymbol #1}}
\newcommand{\Rc}{R_{\rm c}}
\newcommand{\zc}{z_{\rm c}}
\newcommand{\zu}{z_{\rm u}}
\newcommand{\tildezu}{\tilde z_{\rm u}}
\newcommand{\ep}{\epsilon}
\newcommand{\Dt}{\Delta t}
\newcommand{\nus}{\nu_*}
\newcommand{\dc}{d_{\rm c}}
\newcommand{\tauw}{\tau_{\rm w}}

\begin{document}

% Journal identifier can be put here if required, e.g.
%\jl{14}

\title
[ Dynamics of $k$-core percolation in a random graph]
{ Dynamics of $k$-core percolation in a random graph}

\author{Mami Iwata and Shin-ichi Sasa
\footnote[3]
{To whom correspondence should be addressed
(sasa@jiro.c.u-tokyo.ac.jp).} }

\address{Department of Pure and Applied Sciences, University of Tokyo, 
Komaba, Tokyo 153-8902, Japan}

\begin{abstract}
We study the edge deletion process of random graphs near 
a $k$-core percolation point. We find that the time-dependent 
number of edges in the process exhibits  critically divergent 
fluctuations.
We first show theoretically that the $k$-core percolation point 
is exactly given as the saddle-node bifurcation point in a 
dynamical system. We then determine 
all the exponents for the divergence based on a  universal 
description of fluctuations near the saddle-node bifurcation. 
\end{abstract}

\pacs{05.10.Gg,05.70,Jk, 64.60.ah}

%\submitted

%%%%%%%%%%%%%% main text starts %%%%%%%%%%%%%%%%%%%

%%%%%%%%%%%%%%%%%%%%%%%%%
\section{Introduction}  %
%%%%%%%%%%%%%%%%%%%%%%%%%

% k-core

\begin{figure}[htbp]
\begin{center}
\begin{tabular}{ccc}
\includegraphics[width=4cm]{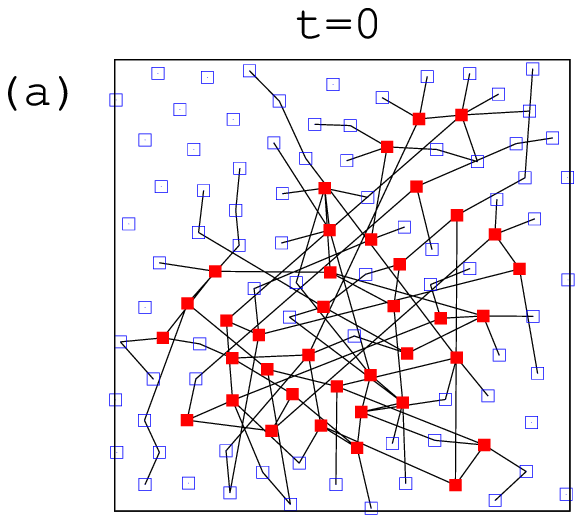}
\includegraphics[width=4cm]{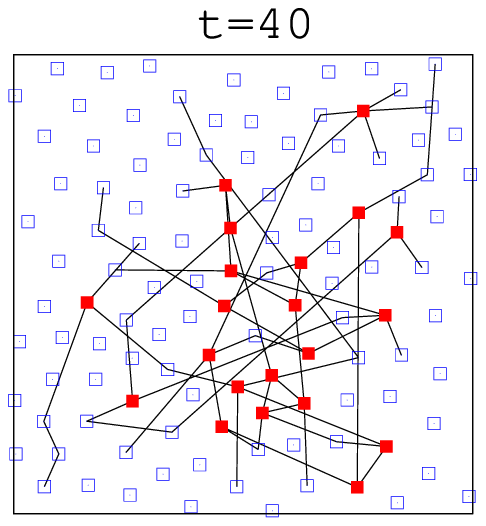}
\includegraphics[width=4cm]{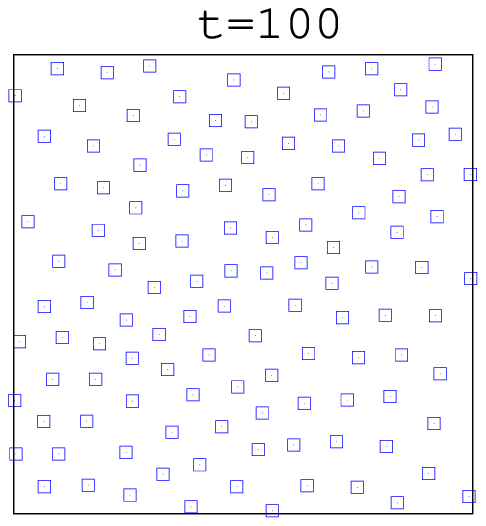}
\end{tabular}
%\caption{..}
\begin{tabular}{ccc}
\includegraphics[width=4cm]{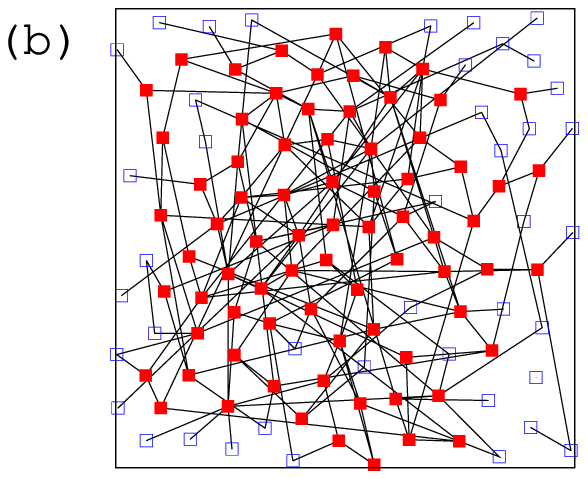}
\includegraphics[width=4cm]{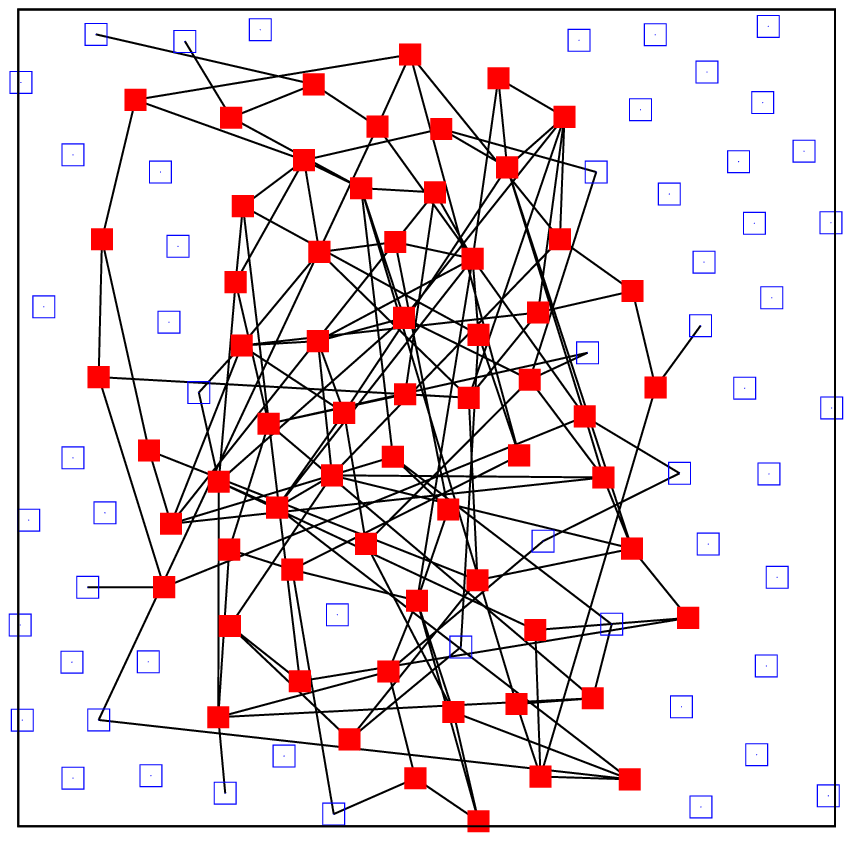}
\includegraphics[width=4cm]{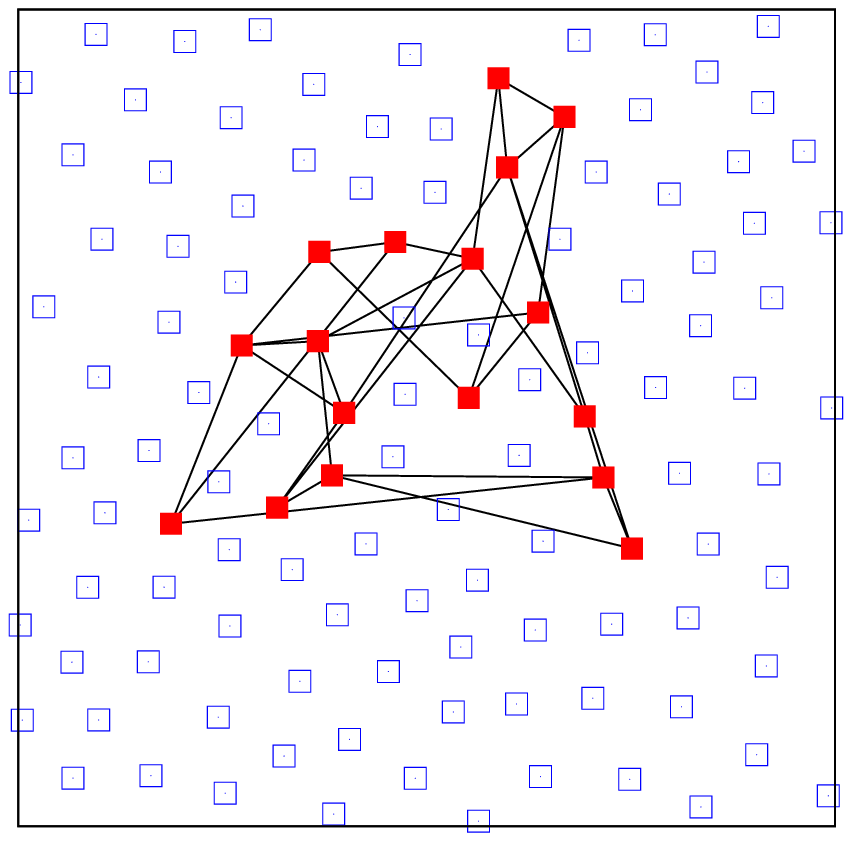}
\end{tabular}
\end{center}
\caption{
Time evolution of random graphs from
initial states (left) to final states (right). $k=3$, $n=118$,
and $R=1.0$ (a) and $R=1.7$ (b), respectively. 
The final state in the case $R=1.7$ corresponds to the $3$-core. 
The filled square symbol (red online) at each time represents 
a heavy vertex and  open square symbol (blue online) a light vertex.}
\label{anime}
\end{figure}

We study the following time evolution of  random graphs on $n$ 
vertices. Let us  denote one sample trajectory of graphs by
$\{ G(t) \}_{0 \le t \le \infty }$, where the time $t$ is a 
real number. 
$G(0)$ is assumed to contain $m$ edges that connect two vertices
chosen randomly. 
Here, $R=m/n$ is regarded as a control parameter of the model. 
When $t>0$, a vertex is chosen with a constant rate $\alpha$ 
for each vertex. Then,  if the number is less than an integer 
$k$, all the edges incident to the vertex are deleted. 
This rule defines the Poisson jump process $G(t)$ in the set of  
graphs on $n$ vertices. We display two examples of the time
evolution of graphs in Fig. \ref{anime}, where the random graphs 
are embedded in the two-dimensional space.

% k-core percolation 

Let $\mu(t)$ be the number of edges at time $t$.
Obviously, when $R$ is sufficiently small, $\mu(t=\infty)/n$ 
is zero with probability 1 in the limit $n \to \infty$. 
It has been known that there is a critical value of $R$ 
above which  $\mu(\infty)/n$ is finite (nonzero)  with 
probability 1 in the same limit \cite{Pittel}. 
The final graph provides the $k$-core (See Fig. \ref{anime}), 
which is defined by the largest subgraph with minimum degree 
at least $k$. This  transition with respect to 
the change in $R$ is called the  $k$-core percolation in a random 
graph. The critical value $\Rc$ was calculated exactly in Ref. 
\cite{Pittel}.

% background of k-core percolation 

The $k$-core percolation was studied in 
several research fields such as magnetism \cite{Reich}, rigidity 
percolation \cite{Leath}, jamming transitions \cite{Chayes,nagel}, 
and network problems \cite{Menders1,Menders2}.
It is also related to the random field Ising 
model, which is a representative model exhibiting so-called 
avalanches \cite{Duxbury2,Dhar}. 
In particular, the dynamics of $k$-core percolation might be 
considered from the viewpoint of the vulnerability of a  
network to random node attack \cite{Duxbury}. 

% our motivation 

\begin{figure}[htbp]
\begin{center}
\begin{tabular}{cc}
\includegraphics[width=6cm]{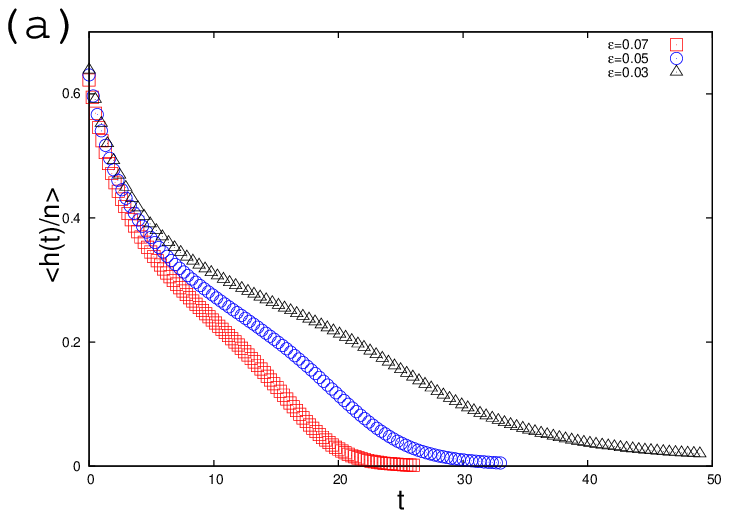}
\includegraphics[width=6cm]{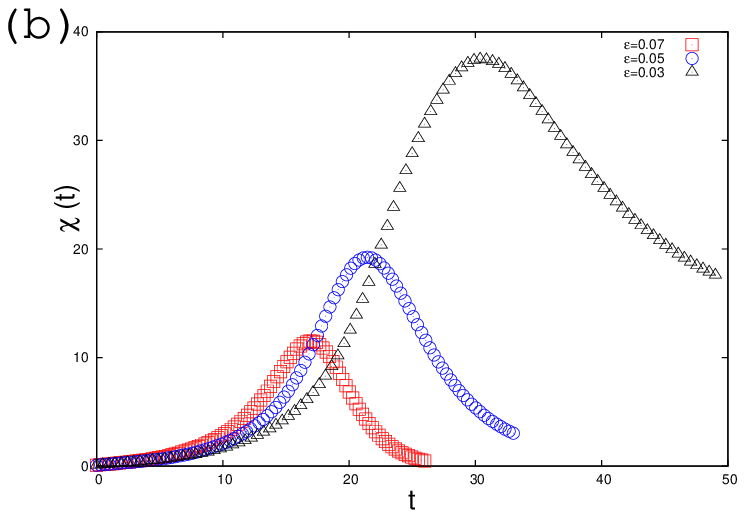}
\end{tabular}
\end{center}
\caption{Relaxation behavior of  heavy vertex density $ \bra h(t)/n \ket $ 
(a) and its fluctuation intensity $\chi(t)$ (b). 
$\epsilon=0.03$, $\epsilon=0.05$, and $\epsilon=0.07$.  
$k=3$, $\alpha=1$, and $n=4096$.}  
\label{fig1}
\end{figure}

In this paper, we wish to elucidate the nature of the dynamics 
near the transition point. 
Concretely, let $h$ be the number of vertices with degree at 
least $k$. (Such a vertex is called a {\it heavy vertex};
otherwise, a {\it light vertex}.) We are interested in the time 
evolution of $h$. As an example, we present the results of 
numerical simulations in Fig. \ref{fig1}.
\footnote{In numerical simulations, we generate a chain of waiting time 
obeying the Poisson distribution and  choose a vertex randomly 
with these time intervals.}
Here, the ensemble average $\bra h(t)/n \ket $ and its fluctuation
intensity 
$\chi(t)\equiv \bra (h(t)-\bra h(t) \ket)^2/n  \ket $ are displayed as 
functions of $t$.  Figure \ref{fig1} indicates that  $\chi(t)$ has 
one peak at $t=\tau$, and we conjecture that $\tau$ and $\chi(\tau)$ 
exhibit the power-law divergences $\tau \simeq \epsilon^{-\zeta}$ and 
$\chi(\tau) \simeq \epsilon^{-\gamma}$, where $\epsilon=\Rc-R > 0$. 
Indeed, we will derive these divergences theoretically and 
determine the values $\zeta=1/2$
and $\gamma=5/2$.

% connection to jamming systems

The divergent behavior observed near the percolation point 
suggests the existence of critical fluctuations. On the other hand, 
it has been known that  a giant $k$-core appears 
in the discontinuous manner at the transition point for  cases $k \ge 3$. 
Such coexistence 
of the discontinuous transition and critical fluctuations has been
emphasized in  relation to the nature of jamming and glass transitions
\cite{BB, Toninelli, MCT}.
Therefore, the theoretical description of the divergent
behavior near the $k$-core percolation may provide a new insight 
toward understanding of jamming and glassy systems.

% road map

This paper is organized as follows. In Sec. \ref{Master},
we present a master equation for four variables that
characterize a graph. Since this master equation 
was derived rigorously in Ref. \cite{Pittel}, our 
presentation in this paper is based on an intuitive 
argument understandable for physicists. Then, 
in Sec. \ref{Langevin}, by considering the situation 
with large $n$, we derive a Langevin equation 
for the four variables. The Langevin
equation is analyzed in the subsequent two sections.
In Sec. \ref{Deteq}, we 
find a saddle-node bifurcation for the rate equation 
obtained by the limit $n \to \infty$. Here, the bifurcation 
point corresponds to the $k$-core percolation point.
Then, in Sec. \ref{Critical},  we study  effects of 
noise near the bifurcation point and calculate 
the exponents that characterize critical divergences.
The final section is devoted to concluding remarks.
In order to simplify the argument, we consider the case $k=3$.
The generalization to cases $k \ge 3$ is straightforward, 
and essentially the same results are obtained. 

%%%%%%%%%%%%%%%%%%%%%%%%%%%%%%%%%%%%%%%%%%%%
\section{Master equation}\label{Master}   %
%%%%%%%%%%%%%%%%%%%%%%%%%%%%%%%%%%%%%%%%%%%%

Let $\Dt$ be a sufficiently small time interval. We can 
describe the stochastic process by the transition probability 
$P(G'|G)$, which is the probability that  $G(t+\Dt)=G'$ 
under the condition that $G(t)=G$. Since $P(G'|G)$ is a 
huge matrix, we cannot treat it directly. 
Hence, we wish to 
have a simple description of the dynamics. 
The simplification of the dynamics of $G(t)$ consists of two 
steps. In the first step, we describe  the dynamics in terms 
of the characteristic quantities of the graph such as 
the number of edges $\mu$ and 
the number of vertices $v_r$ with degree $r$, where 
$r=0,1,\cdots$.  Among them, the number of light vertices, $v_0$,
$v_1$ and $v_2$ are directly related to the dynamics of 
the graph because all the edges incident to a chosen light vertex  
will be deleted in the next change of the graph.
Indeed, according to Ref. \cite{Pittel},
\footnote{ 
Note that the edge deletion process in Ref. \cite{Pittel}
is not identical to the dynamics we define. 
First, their dynamics are given 
as a discrete process. Second, in their dynamics, 
a non-isolated light vertex is always chosen at each time 
step. Thus, as time goes on, the deletion in their dynamics
is accelerated more than that in our evolution rule.
Despite this difference, one can transform mathematical 
statements in \cite{Pittel} to those valid in our model. }
the time evolution of the four-tuple  $\bv{w}=(\mu, v_0,v_1,v_{2})$ 
is described by a Markov process. Mathematically,  
the probability of $\bv{w}'$ at time $t+\Delta t$ provided 
that $\bv{w}$ at time $t$ is given, which is denoted by 
$p(\bv{w'}|\bv{w})$, is  expressed as a function of $\bv{w'}$ 
and $\bv{w}$ for general $n$. (See proposition 1 in 
Ref. \cite{Pittel}.) Subsequently, in the second step of 
the simplification, the asymptotic formula of $p(\bv{w'}|\bv{w})$ 
for large $n$ is derived. (See Corollary 1 in Ref. \cite{Pittel}.)

In this paper, we do not review the derivation of the asymptotic 
form $p(\bv{w'}|\bv{w})$ in Ref. \cite{Pittel}. Instead, we provide 
its mathematically naive  derivation by focusing  on cases with 
large $n$ from the outset. More precisely, we estimate $p(\bv{w'}|\bv{w})$
assuming that $v_r/n$ with $r \ge 3$ takes the most probable
value $h q_r /n$, where  from the law of large numbers, $q_r$ is equal 
to the probability that $r$ edges are incident to a given heavy 
vertex under the condition that $\bv{w}$ is specified. (Recall that $h$ 
represents the number of heavy vertices that is equal to 
$\sum_{r \ge k=3} v_r$. ) We express  this statement formally as 
\begin{equation}
\frac{v_r}{h}= q_r,
\label{v3}
\end{equation}
where the probability $q_r$ is given by the Poisson distribution 
\begin{equation}
q_r = \frac{1}{r!} \frac{1}{Q(z)}z^r e^{-z}
\label{qjdef}
\end{equation}
with the normalization constant
\begin{eqnarray}
Q(z)  &=& \e^{-z}(e^{z}-1-z-z^2/2).
\label{Qdef}
\end{eqnarray}
Although the appearance of the Poisson distribution seems
natural, its mathematical proof is not simple. 
(See Ref. \cite{Pittel}.)  Here, using the trivial 
relation\footnote{
We express an edge by a pair of integers
$(\ell_1, \ell_2)$ when the edge links 
a vertex of degree  $\ell_1$ and another
vertex of degree $\ell_2$.
Collecting all the pairs of integers for the $\mu$ 
edges, we have $2\mu$ integers. Here, count the
number of integers that are greater than $2$. }
$2\mu-v_1-2v_2=\sum_{r=3}^\infty  r v_r $,
we have 
\begin{equation}
\frac{2\mu-v_1-2v_2}{h}=\sum_{r=3}^\infty  r q_r ,
\label{zdet}
\end{equation}
which determines $z$ in (\ref{qjdef}) and (\ref{Qdef})
for a given $\bv{w}$.  
Since the direct calculation using (\ref{qjdef})
leads to 
\begin{equation}
\sum_{r=3}^\infty r q_r=\frac{z \Pi(z)}{Q(z)}
\label{rvr4}
\end{equation}
with
\begin{equation}
\Pi(z)=\e^{-z}(e^z-1-z),
\end{equation}
we obtain a useful relation for the determination of $z$ from $\bv{w}$:
\begin{eqnarray}
\frac{2\mu-v_1-2v_2}{h}&=&
\frac{z\Pi(z)}{Q(z)}.
\label{zdet2}
\end{eqnarray}
In the argument below, $z$ always represents the unique solution of
(\ref{zdet2}) for a given $\bv{w}$.  Furthermore, one can 
easily confirm the relation 
\begin{equation}
z=\sum_{r\ge 4} r q_r,
\label{zmean}
\end{equation}
which provides us a simple interpretation of $z$.

%%% idea 

Now, we estimate $p(\bv{w'}|\bv{w})$.
We first notice the value of $\mu'-\mu$. (Note that
$\mu'-\mu$ represents the change of the number of edges
during the time interval $\Delta t$.)
(i) When $\mu'-\mu=0$,  no deletion occurs. 
This implies $\bv{w'}=\bv{w}$.
(ii) When $\mu'-\mu=-1$,  one edge incident to a chosen
vertex is deleted. This edge connects the chosen vertex with
another vertex with degree $\ell$. Then, $\bv{w'}-\bv{w}$ 
takes four values depending on $\ell=1, 2,3$,  and $ \ell \ge 4$,
which are denoted by $\bv{\sigma}_j$ with $j=1,2,3$ and
$\bv{\sigma}_4$, respectively. (See Table. \ref{table1}.) 
(iii) When $\mu'-\mu=-2$,  two edges incident to 
a chosen vertex are  deleted. 
Each  edge connects the chosen vertex to another vertex  with 
degree $\ell_i$, $i=1,2$, where $\ell_i=1,2,3$ or $\ell_i \ge 4$. 
We assume $\ell_1 \le \ell_2$
without loss of generality. Then, $\bv{w'}-\bv{w}$ takes 
ten  values depending on the values 
of $\ell_1$ and $\ell_2$, 
which are denoted by $\bv{\sigma}_j$, $5 \le j  \le 14$,
where the correspondence between $j$ and $(\ell_1, \ell_2)$
is shown in Table \ref{table2}. (iv) We do not need to consider the cases 
$\mu'-\mu \le -3  $. Although such cases appear when 
deletions occur twice or more  during the time interval,
the probability of their occurrence is negligible for 
sufficiently small $\Dt$.
To sum up,  $\bv{w'}-\bv{w}$ takes 
either $\bv{0}$ or  $\bv{\sigma}_j$, $j=1,\cdots, 14$,
and the fourteen transitions occur independently. 

% rate

We denote the rate of  transition $\bv{w} \to \bv{w}+\bv{\sigma}$ by 
$r(\bv{\sigma}|\bv{w})$. We can then write 
\begin{equation}
p(\bv{w'}|\bv{w})= n \Delta t \sum_{j=1}^{14}  r (\bv{\sigma}_j|\bv{w})
\delta(\bv{\sigma}_j, \bv{w'}-\bv{w})
\label{transitionprob_wwd}
\end{equation}
when $\bv{w'} \not = \bv{w}$. $\delta(\bv{x},\bv{y})$ is the
four-dimensional Kronecker delta function for $\bv{x}$, $\bv{y}$
in ${\bf N}^4$. $p(\bv{w}|\bv{w})$ is
determined from the normalization condition of the
probability. 

\begin{table}
\begin{center}
\begin{tabular}{c|c|c}
\hline
$j$  &  $\bv{\sigma}$  & $r(\bv{\sigma}|\bv{w})$  \\
\hline
1 & (-1,2,-2,0) & $\alpha v_1 p_1/n$ \\
2 & (-1,1,0,-1) & $\alpha v_1 p_2/n$ \\
3 & (-1,2,-1,1) & $\alpha v_1 p_3/n$ \\
4 & (-1,1,-1,0) & $\alpha v_1 p_4/n$ \\
\hline
\end{tabular} 
\caption{Rate $r(\bv{\sigma_j}|\bv{w})$ for the
transition $\bv{w} \to \bv{w}+\bv{\sigma_j}$.
$p_j = jv_j/(2\mu)$ for $1 \le j \le 3$, 
and $p_4=zh/(2\mu)$.}
\label{table1}
\end{center}
\end{table}

% estimation of the rate

Let us estimate the transition rate $r(\bv{\sigma}_j|\bv{w})$.
Recall that a vertex is randomly chosen at the rate $\alpha$ 
(per unit time and per each vertex). Then, the probability that 
the degree of the chosen vertex equals to one is given by 
$v_1/n$. We next 
consider the conditional probability that the edge incident
to the chosen vertex connects it with a vertex of degree $r$.
This probability, which is denoted by $c_r$, takes a complicated
form for general cases. (See  Ref. \cite{Pittel}). 
Here, notice that $r v_r/2\mu $ is the probability of finding 
a vertex with degree $r$ when  we observe one vertex connected 
to an edge, which we choose randomly. The difference between 
$c_r$ and $r v_r/2\mu$ originates from the condition under which 
an edge is chosen.  The difference 
is negligible for sufficiency large $n$. 
Combining these results,  the rate 
$r(\bv{\sigma}_j|\bv{w})$ 
is estimated as 
\begin{equation}
r(\bv{\sigma}_j|\bv{w})= \alpha \frac{v_1}{n} \frac{j v_j}{2 \mu}
\label{rate13}
\end{equation}
for $1 \le j \le 3$, and 
\begin{equation}
r(\bv{\sigma}_4|\bv{w})= \alpha \frac{v_1}{n} \sum_{r=4}^\infty 
\frac{r v_r}{2 \mu}.
\label{rate4}
\end{equation}
Here, it should be noted that 
$ v_3$ and  $\sum_{r=4}^\infty r v_r$ in (\ref{rate13}) and 
(\ref{rate4}) are calculated from  (\ref{v3}) with $z$
determined by (\ref{zdet2}). For convenience 
of later calculation, we summarize the result in Table \ref{table1}. 
In this table, we introduce $p_j = jv_j/(2\mu)$ for $1 \le j \le 3$, 
and $p_4=\sum_{r=4}^\infty r v_r/(2\mu)=zh/(2\mu)$. 
(We used (\ref{zmean}) in deriving the latter equality.)  
The transition rate $r(\bv{\sigma}_j|\bv{w})$ with $5 \le j \le 14$
is calculated in the same manner by noting that two edges incident to one
vertex can be treated 
independently. The result is summarized in Table \ref{table2}.
In the argument below, we set $\alpha=1$ for simplicity.

\begin{table}
\begin{center}
\begin{tabular}{c|c|c|c}
\hline
$j$  & $(\ell_1,\ell_2)$ &  $\bv{\sigma}$  & $r(\bv{\sigma}|\bv{w})$  \\
\hline
5 & (1,1)  &  (-2,3,-2,-1) & $\alpha v_2 p_1^2/n$ \\
6 & (1,2)  &  (-2,2,0,-2) & $2 \alpha v_2 p_1p_2/n$ \\
7 & (1,3)  &  (-2,2,-1,0) & $2 \alpha v_2 p_1p_3/n$ \\
8 & $(1,\ge 4)$  &  (-2,2,-1,-1) & $2 \alpha v_2 p_1 p_4/n$ \\
9 & (2,2)  &  (-2,1,2,-3) & $\alpha v_2 p_2^2/n$ \\
10 & (2,3)  &  (-2,1,1,-1) & $2\alpha v_2 p_2p_3/n$ \\
11 & $(2,\ge 4)$  &  (-2,1,1,-2) & $2\alpha v_2 p_2p_4/n$ \\
12 & (3,3)  &  (-2,1,0,1) & $\alpha v_2 p_3^2/n$ \\
13 & $(3, \ge 4)$  &  (-2,1,0,0) & $2\alpha v_2 p_3p_4/n$ \\
14 & $(\ge 4, \ge 4)$  &  (-2,1,0,-1) & $\alpha v_2 p_4^2/n$ \\
\hline
\end{tabular} 
\caption{
Rate $r(\bv{\sigma_j}|\bv{w})$ for the
transition $\bv{w} \to \bv{w}+\bv{\sigma_j}$.
$p_j = jv_j/(2\mu)$ for $1 \le j \le 3$, 
and $p_4=zh/(2\mu)$.}
\label{table2}
\end{center}
\end{table}

%%%% initial condition 

Before closing this section, we consider the initial 
condition $\bv{w(0)}$. 
In order to simplify the argument, we assume that 
$\bv{w(0)}/n$ takes the most probable value in the limit 
$n \to \infty$. Let us calculate this value. We first 
consider the probability that $k$ edges are incident to
a vertex chosen randomly: 
\begin{eqnarray}
P_k &=&
\frac{{}_{(n-1)} C _k \cdot {}_{(N-n+1)} C _{(m-k)}}{{}_N C _m},
\end{eqnarray}
where $N={}_n {\rm C}_2$.  
Taking the limit $n \to \infty$ with  fixing $R(=m/n)$,  we
obtain 
\begin{eqnarray}
P_k &=&\frac{1}{k!}(2R)^ke^{-2R}.
\end{eqnarray}
This leads to 
\begin{equation}
\bv{w(0)}=(m, n e^{-2R}, n (2R) e^{-2R},n (2R)^2e^{-2R}/2).
\label{init}
\end{equation}

%%%%%%%%%%%%%%%%%%%%%%%%%%%%%%%%%%%%%%%%%%%%%%%
\section{Langevin equation}\label{Langevin}   %
%%%%%%%%%%%%%%%%%%%%%%%%%%%%%%%%%%%%%%%%%%%%%%%

We define the density variable 
$\bv{\rho}=(\bar \mu, \rho_0,\rho_1,\rho_2)$ by
$\bv{w}/n$. When $n$ is large, the dynamics of $\bv{\rho}$ are
expected to be 
described by a Langevin equation. We shall derive the equation from 
the transition probability $p(\bv{w}'|\bv{w})$ given
in (\ref{transitionprob_wwd}) with Tables \ref{table1} and \ref{table2}. 

We utilize an  expansion formula of the Kronecker delta function
\begin{equation}
\delta(x,y)=\int_{-\pi}^\pi  \frac{du}{2 \pi} \e^{i u (x-y)}
\end{equation}
for any $x$ and $y$ in ${\bf N}$. Substituting this formula into 
(\ref{transitionprob_wwd}), we obtain 
\begin{equation}
p(n \bv{\rho'}| n \bv{\rho})= n \Delta t \int_{\cal D}  
\frac{d^4\bv{u}}{(2 \pi n)^4} 
\e^{-i \bv{u}  \cdot (\bv{\rho'}-\bv{\rho})} 
\sum_{j=1}^{14}   r (\bv{\sigma}_j|n \bv{\rho})
\e^{i \bv{u}\cdot \bv{\sigma_j}/n}
\end{equation}
for $\bv{\rho}' \not = \bv{\rho}$, where  ${\cal D}=[-n\pi, n\pi]^4$.
Noting $p(n \bv{\rho}| n \bv{\rho})=
1-\sum_{j=1}^{14} r (\bv{\sigma}_j|\bv{w})n\Dt$, we can write
\begin{equation}
p(n \bv{\rho'}| n \bv{\rho})= \int_{\cal D}  
\frac{d^4\bv{u}}{(2 \pi n)^4} \e^{-i \bv{u}  \cdot (\bv{\rho'}-\bv{\rho})} 
{\cal F}(\bv{\rho})
\label{Fdef}
\end{equation}
for any $\bv{\rho'}$ and $\bv{\rho}$, where ${\cal F}$ is 
expressed as 
\begin{eqnarray}
{\cal F}(\bv{\rho}) 
&=&
n\Dt\sum_{j=1}^{14}   r (\bv{\sigma}_j|n \bv{\rho})
\e^{i \bv{u}\cdot \bv{\sigma_j}/n}
+\lr{1
-\sum_{j=1}^{14}   r (\bv{\sigma}_j|n \bv{\rho})n\Dt
}\nm
&=& 
\exp
\lr{
\sum_{j=1}^{14}   r (\bv{\sigma}_j|n \bv{\rho})n\Dt
\lr{
\e^{i \bv{u}\cdot \bv{\sigma_j}/n}-1
} + O(\Dt^2)} \nm
&\simeq&
\exp
\lr{
\sum_{j=1}^{14}   r (\bv{\sigma}_j|n \bv{\rho})
n\Dt
\lr{ 
i \bv{u}\cdot \bv{\sigma_j}/n- 
(\bv{u}\cdot \bv{\sigma_j})^2/(2n^2)
} },
\label{exp_dt}
\end{eqnarray}
where we have ignored the terms of $O( (\Delta t)^2, \Delta t/n^2)$.
Here, we first consider the case
with large $n$ and then assume $\Delta t$ to be sufficiently small.
We then have the transition probability 
\begin{eqnarray}
p(n \bv{\rho'}|  n \bv{\rho})
&=& \int_{{\bf R}^4}  \frac{d^4\bv{u}}{(2 \pi n)^4} 
\exp\left[
-i{\bv{u}}\cdot\lr{
\bv{\rho'}-\bv{\rho} -\sum_{j=1}^{14}  
r (\bv{\sigma}_j|n \bv{\rho}) \Dt\bv{\sigma_j}
} \right. \nm
& & - \left.
\sum_{j=1}^{14}   r (\bv{\sigma}_j|\bv{n \rho})\Dt
\lr{{ \bv{u}} \cdot\bv{\sigma_j}}^2\frac{1}{2n}
\right].
\label{pww}
\end{eqnarray}
Here, we define a $4 \times 4$ matrix 
\begin{eqnarray}
A_{lm}(n\bv{\rho})
\equiv
\sum_{j=1}^{14}
\frac{r (\bv{\sigma}_j|n \bv{\rho})}{2}
(\bv{\sigma}_j)_l(\bv{\sigma}_j)_m.
\end{eqnarray}
Since the matrix $\hat A=(A_{lm})$ is semi-positive, there exists
the semi-positive matrix $\hat G$ satisfying 
$\hat A= \hat G^2$.
We also define $\Delta \bv{\Xi}$ by 
\begin{eqnarray}
\hat G(n \bv{\rho})  \Delta \bv{\Xi}\equiv
{\bv{\rho'}}-{\bv{\rho}}
-\sum_{j=1}^{14}r (\bv{\sigma}_j|n\bv{\rho}) 
\Dt \bv{\sigma_j}.
\label{xi_def}
\end{eqnarray}
Then, the probability density of $\Delta \bv{\Xi}$ 
is expressed as
\begin{equation}
p(\Delta \bv{\Xi})=\det(\hat G)n^4p(n\bv{\rho'}|n\bv{\rho}),
\label{pdef}
\end{equation}
where $\det(\hat G)$ is the determinant of the Jacobian matrix
associated with the transformation from $\bv{\rho'}$
to $\Delta \bv{\Xi}$, and note that the probability density 
of $\bv{\rho'}$ 
is given by $n^4p(n\bv{\rho'}|n\bv{\rho})$, because 
$\int d^4 \bv{\rho'} p(n\bv{\rho'}|n\bv{\rho})=
\sum_{\bv{w'}} p(\bv{w'}|n\bv{\rho})/n^4 =1/n^4$.
From (\ref{pww}) and (\ref{pdef}), we obtain
\begin{eqnarray}
p(\Delta \bv{\Xi})
&=&
\frac{n^2}{16 \pi^2 (\Delta t)^2 } 
\e^{-\frac{n}{4 \Delta t}  
(\Delta\bv{\Xi}) (\Delta\bv{\Xi})}.
\label{noise_bunpu}
\end{eqnarray}
This implies that $\Delta \bv{\Xi}$ is the Gaussian noise
satisfying
\begin{eqnarray}
\bra \Delta\Xi_l\Delta\Xi_m\ket
&\equiv&\frac{2\Dt}{n }\delta_{lm}.
\label{xi_amp}
\end{eqnarray}
Taking the limit $\Dt\to 0$ in (\ref{xi_def}) with (\ref{xi_amp}),
we obtain
\begin{eqnarray}
\partial_t {\bv{\rho}}
&=&
\sum_{j=1}^{14}r (\bv{\sigma}_j|n \bv{\rho})  \bv{\sigma_j}+
\sqrt{\frac{1}{n}}\hat G(n\bv{\rho}) \cdot\bv{\xi},
\label{langevin}
\end{eqnarray}
with $\bra\xi_l(t)\xi_m(t')\ket=2\delta(t-t')\delta_{lm}$.
Here, the symbol $'' \ \cdot \ ''$ in (\ref{langevin}) represents the Ito rule
of the multiplication of stochastic variables.
% initial condition
Finally, from (\ref{init}), the initial condition of the 
Langevin equation  is given by
\begin{equation}
\bv{\rho(0)}=(R,  e^{-2R},  (2R) e^{-2R}, (2R)^2e^{-2R}/2).
\label{Lan-init}
\end{equation}

%%%%%%%%%%%%%%%%%%%%%%%%%%%%%%%%%%%%%%%%%%%%%%%%
\section{Deterministic equation}\label{Deteq}  %
%%%%%%%%%%%%%%%%%%%%%%%%%%%%%%%%%%%%%%%%%%%%%%%%

The Langevin equation (\ref{langevin}) becomes the deterministic equation
in the limit $n \to \infty$: 
\begin{eqnarray}
\partial_t \bv{\rho}
&=&\sum_{j=1}^{14}r(\bv{\sigma}_j| n\bv{\rho})  \bv{\sigma_j}.
\label{rate_1}
\end{eqnarray}
Concretely, using Tables \ref{table1} and \ref{table2}, 
we can obtain the expression 
\begin{eqnarray}
\partial_t \bar \mu&=&-s,
\label{odem} \\
\partial_t \rho_0 &= & \rho_1+\rho_2
+\rho_1 \frac{s}{2 \bar \mu}, 
\label{ode0}\\
\partial_t \rho_1 &=& -\rho_1-\rho _1 \frac{s}{2\bar \mu}
+2 \rho_2 \frac{s}{2\bar  \mu},
\label{ode1}\\
\partial_t \rho_2 &=& -\rho_2-2 \rho_2 \frac{s}{2\bar \mu}+ 
3 \rho_3\frac{s}{2\bar  \mu},
\label{ode2}
\end{eqnarray}
where $s=\rho_1+2 \rho_2$ is the density  of the edges incident to 
light vertices, and $\rho_3=v_3/n$ is determined as a function of
$\bv{\rho}$ from (\ref{v3}) with $z$ determined by
(\ref{zdet2}).  
The derivation of (\ref{odem})-(\ref{ode2}) requires tedious 
calculation, while the result is understood intuitively. 
For example, the third term of (\ref{ode0}) represents
the change in the degree of a vertex from 1 to 0 by the deletion
of the edge connecting this vertex with another vertex that is chosen
randomly.
The differential equation given in (\ref{odem}) - (\ref{ode2})
with the initial condition (\ref{Lan-init}) determines the most probable
behavior of the Langevin equation (\ref{langevin}). 

% transformation of variable

Putting aside the initial condition, we study the differential
equation in (\ref{odem}) - (\ref{ode2}). The complicated nature
arises from the implicit dependence of $\rho_3$ on $\bv{\rho}$ 
through $z$. In order to avoid it, we
carry out the transformation of variables. First, we choose $z$ as
a dynamical variable. Taking the derivative of (\ref{zdet2})
with respect to time, we obtain
\begin{equation}
\partial_t z= -\frac{sz}{2\bar \mu}.
\label{odez}
\end{equation} 
Seeing  (\ref{odem}) and (\ref{odez}), we further choose
$s$ as a dynamical variable. We then obtain
\begin{equation}
\partial_t s= -s-\frac{s^2}{2 \bar \mu}+3\frac{v_3 s}{\bar \mu}.
\label{odes}
\end{equation}
Thus, (\ref{odem}), (\ref{ode1}), (\ref{odez}), and (\ref{odes})
constitute the differential equation for 
$(\bar \mu, \rho_1, s, z)$, which is equivalent to the
differential equation for $\bv{\rho}$. 
Here, from (\ref{odem}) and (\ref{odez}), we find
immediately a constant of motion
\begin{equation}
J_1=\frac{z^2}{\bar \mu}.
\end{equation}
Furthermore, noting $\partial_t  \bar h=-3 \rho_3 s/(2\bar \mu)$
and $\partial_z Q=z^2 \e^{-z}/2$, one confirms that 
there is another constant of motion 
\begin{equation}
J_2=\frac{\bar h}{Q(z)},
\label{35}
\end{equation}
where $\bar h=h/n$. 
Recalling (\ref{zdet2}), we rewrite $J_2$ as 
$J_2= z \Pi(z)/(2 \bar \mu-s)$. 
Now, defining $J_3=J_1 J_2/2$, we obtain the expression
\begin{equation}
s= 2 \left( 1- J_3 \frac{\Pi(z)}{z} \right) \frac{z^2}{J_1},
\label{sint}
\end{equation}
which  defines integral curves in the $(z,s)$ space.

% normal form 

Using the constants of motion, we choose a set of 
dynamical variables as $ \bv{\zeta}=(J_1, J_3, \rho_1, z)$. 
Then, the differential equation for $(J_1, J_3, z)$
takes the simplest expression that $\partial_t J_1 = 0$, 
$\partial_t J_3 = 0$, and 
\begin{equation}
\partial_t z = -z + J_3 \Pi(z).
\label{zeq1}
\end{equation}
Importantly, the time evolution of $z$ is independent of $\rho_1$. 
Besides, the initial condition (\ref{Lan-init}) leads to 
$J_1=4R$ and $J_2=1$; thus, $J_3=2R$. To sum up,
the dynamical behavior of the edge deletion process 
of random graphs is described by 
\begin{equation}
\partial_t z =-z +2 R \Pi(z)
\label{zeqfinal}
\end{equation}
with $z(0)=2R$. 

% critical point

\begin{figure}[htbp]
\begin{center}
\begin{tabular}{ccc}
\includegraphics[width=6cm]{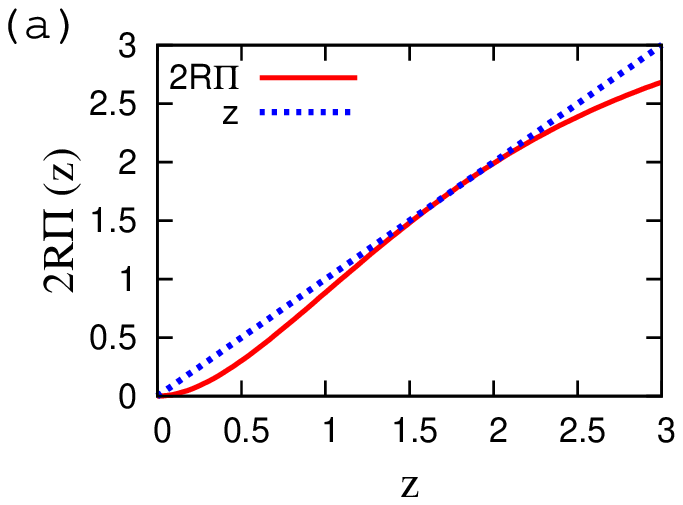}
\includegraphics[width=6cm]{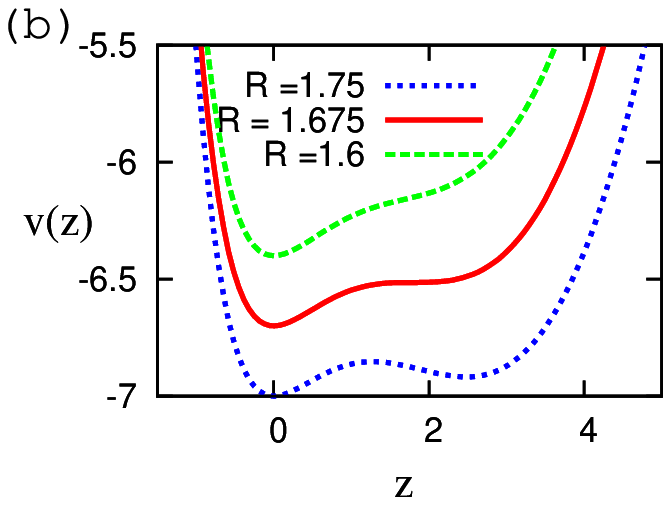}
\end{tabular}
\end{center}
\caption{(a) The graphs of $z$ and $2R \Pi(z)$ with $R=1.675$.
Two intersection points appear when $R$ is increased, 
while no intersection when $R$ is decreased. 
(b) Shapes of potential functions $V(z)$ for the cases
that $R \simeq \Rc$, $R < \Rc$ and $R > \Rc$, respectively.}
\label{fig0}
\end{figure} 

The differential equation given in (\ref{zeqfinal}) 
can be easily analyzed. First, there exists the trivial solution 
$z=0$. Then, let $z_*$ be another fixed point (if it exists). 
$z_* (\not = 0) $ satisfies $z_*=2R \Pi(z_*)$. From  Fig. \ref{fig0}, 
we find that  two nontrivial solutions exist when $ R > \Rc$, where
$\Rc$ is determined by $\min_{z >0} [z- 2\Rc \Pi(z)]=0$, which 
yields the compact expression of $\Rc$ as
\begin{equation}
\Rc= \frac{1}{2}\min_{z >0 }\left( \frac{z}{\Pi(z)} \right) .
\label{rc}
\end{equation}
We numerically calculated $\Rc \simeq 1.675$. In order to investigate 
the solution trajectory of (\ref{zeq1}), we express  (\ref{zeq1}) 
in the form
\begin{equation}
\partial_t z=-\der{V}{z},
\end{equation}
where the potential function $V(z)$ is given by 
$V(z)=z^2/2-2R(z+2e^{-z}+z e^{-z})$. We display the shape of 
the potential $V(z)$ in Fig. \ref{fig0}.  
It is seen  that there
are a pair of minimum and maximum in addition to
the trivial minimum point $z=0$ when $R > \Rc$. Obviously, the solution 
corresponding to the maximum  (saddle) is unstable, while
the solution corresponding to the minimum  (node) is stable.
Note that the potential is a monotonic increasing
function in $z$ when $R < \Rc$, which corresponds to the 
fact that there is no nontrivial stationary solution when 
$R < \Rc$.  The  qualitative change of trajectories at $R=\Rc$ 
is called a {\it saddle-node bifurcation}. The fixed point at $R=\Rc$ 
is the {\it marginal saddle}, which is denoted 
by $\zc(=z_*(\Rc))$.
Since the condition $z(t \to \infty) \not =0$
represents the existence of a $k$-core, $\Rc$ is the $k$-core percolation
point. This determination method of $R_{\rm c}$ is 
essentially equivalent to that in Ref. \cite{Pittel}.
The achievement of this study is the identification of 
the bifurcation type observed in the dynamics of $k$-core 
percolation in a random graph.

% dynamics; phase / R_c-\epsilon

Now, we investigate the behavior of the system with $R=\Rc-\epsilon$, 
where $\epsilon $ is a small positive constant. We define a dynamical
variable $\phi$ by $z=\zc+\phi$. Substituting it into (\ref{zeqfinal}),
we obtain
\begin{equation}
\partial_t \phi=-\epsilon a  - b  \phi^2  +O(\phi^3),
\label{veq}
\end{equation}
where $a=2 \Pi(\zc)$ and $b=-\Rc \Pi''(\zc)$.
Since the solution $\phi(t)$ is expressed as a scaling form
\begin{equation}
\phi(t)= \epsilon^{1/2} \Phi(\epsilon^{1/2} t),
\label{tscaling}
\end{equation}
the typical time for exiting the marginal saddle is 
proportional to $\epsilon^{-1/2}$.

%%%%%%%%%%%%%%%%%%%%%%%%%%%%%%%%%%%%%%%%%%%%%%%%%%%%%%%%%%%%
\section{Critical fluctuation}\label{Critical}             %
%%%%%%%%%%%%%%%%%%%%%%%%%%%%%%%%%%%%%%%%%%%%%%%%%%%%%%%%%%%%

% normal form of the Langevin equation

Next, we study the  fluctuations that
are described by the Langevin equation (\ref{langevin}).
Since the deterministic equation for 
$\bv{\zeta}$ is the simplest one, we rewrite the
Langevin equation by using $\bv{\zeta}$.  Formally,
we express the variable transformation 
from $\bv{\rho}$ to $\bv{\zeta}$ as 
$ \bv{\zeta} = \bv{\phi} ( \bv{\rho}) $.
Then, using Ito's  formula, one can derive
\begin{equation}
\partial_t z=-z + J_3 \Pi(z)
+ \sum_{jl}\pder{z}{ \rho_j}  \sqrt{\frac{1}{n}} G_{jl} \xi_l  
+ \frac{1}{n} \sum_{jl} \pdert{z}{ \rho_j}{\rho_l} A_{jl},
\label{langevinz}
\end{equation}
where the expression $\partial z/\partial \rho_j$ is evaluated 
from the functional dependence of $z$ on $\bv{\rho}$.
It should be noted that  $J_3$ fluctuates in the Langevin description.  

%% formulation 

Concretely, we investigate the divergent behavior of the 
quantity
\begin{equation}
\chi_z(t)= n(\bra z(t)^2 \ket -\bra z(t) \ket^2) 
\label{chiz}
\end{equation}
for the system with $R=\Rc-\epsilon$, where $\epsilon$ is 
a small positive constant. Here, it is naturally expected that 
the divergent part of the fluctuations of $h(t)$ is identical 
to that of $z(t)$. We therefore conjecture that $\chi_z(t)$
has a peak at $t=t_*$ and that $t_*$ and $\chi_z(t_*)$ exhibit
the power-law divergences $t_* \simeq \ep^{-\zeta}$ and
$\chi_z(t_*)  \simeq \ep^{-\gamma}$. We shall derive 
these divergences theoretically. 

The perturbative calculation with respect to the nonlinearity
in (\ref{langevinz}) seems quite difficult to capture the
divergent behavior of $z(t)$. Instead, we utilize the 
bifurcation structure, as done in  Ref. \cite{Iwata2}. 
Following the idea of the method,  we  first notice  two solutions of 
(\ref{zeqfinal}) with $R=\Rc$. One solution $\zu$ 
satisfies the  conditions 
$\zu(t) \to  \zc$ for $t \to -\infty $, $\zu(t) \to  0$ for 
$t \to \infty $, and $\zu(0)=z_0$, (say, $z_0=0.5)$.
The other solution $z_{\rm B}(t)$ satisfies 
the  conditions $z_{\rm B}(t) \to  \zc$ for $t \to \infty $ 
and $z_{\rm B}(0)=z(0)$. 
$z_{\rm B}(t)$ represents  the down-hill trajectory to 
the marginal saddle from the point $z_0$ in the potential shape.
Then, we express the trajectories by 
using the  exit  time $\theta$ from the marginal saddle as 
\begin{equation}
z(t)=\zu(t-\theta)+(z_{\rm B}(t)-\zc)+\varphi(t-\theta),
\end{equation}
where $\varphi(t-\theta)$ represents a deviation from the superposition
of the two solutions. 

It is worthwhile to note that the variable $\theta$ corresponds
to the Goldstone mode associated with the time-translational symmetry. 
Thus, the fluctuation of $\theta$ carries a divergent part, while  
$\varphi$ can be treated as a variable slaved to $\theta$. 
Based on this observation, $\bra z(t) \ket$ and $\chi_z(t)$ can be
estimated by the statistical average over $\theta$. Note that we have 
devised a theoretical framework in which the statistical distribution 
of $\theta$ can be calculated perturbatively by considering the 
interaction of $\theta$ with $(z_{\rm B}(t)-1)$ \cite{Iwata2, Iwata3}. 
In the argument below, without entering this lengthy calculation,
we shall determine phenomenologically the exponents characterizing 
the divergent behavior of $\chi_z(t)$. 

We first calculate the exponents characterizing the 
divergences of $\bra \theta \ket $ and the intensity
of fluctuation $\chi_\theta$ defined by
\begin{equation}
\chi_\theta \equiv n(\bra \theta^2 \ket -\bra \theta \ket^2).
\end{equation}
We start with the scaling relations 
\begin{eqnarray}
\bra \theta \ket  &=& n^{\zeta'/\nus} f_1( n^{1/\nus} \epsilon) , 
\label{sca1} \\
\chi_\theta  &=& n^{\gamma'/\nus} f_2( n^{1/\nus} \epsilon) ,
\label{sca2} 
\end{eqnarray}
for large $n$ and small $\epsilon$, where 
we have introduced the exponents $\zeta'$, $\nus$, and $\gamma'$. 
We here assume 
that  $f_1(0)={\rm const}$ and $f_2(0)={\rm const}$. We also
assume that $f_1(x) \simeq x^{-\zeta'}$ and $f_2(x) \simeq x^{-\gamma'}$ 
for $x \gg 1$, because $\bra \theta \ket $ and $\chi_\theta$
are expected to be independent of $n$ in the regime $x \gg 1$.
We find that $\zeta'=1/2$  from (\ref{tscaling}).
Furthermore, from (\ref{sca1}), we assume that 
a distribution function of $\theta$ is expressed as an $n$-independent
function of $ \theta n^{-\zeta/\nus}$ when $\epsilon=0$. 
This leads to a relation   $\gamma'/\nus =2 \zeta'/\nus +1$, which 
yields  
\begin{equation}
\gamma'=2 \zeta'+\nus.
\label{rel}
\end{equation}

To this point, we have avoided the analysis of (\ref{langevinz}). 
In order to determine 
the value of $\nus$, we need to study the equation. For the sake of 
a simple argument, we  assume that the behavior near 
the saddle-node bifurcation point is described by (\ref{veq}) with  
a noise term:
\begin{equation}
\partial_t \phi=-\epsilon a  - b  \phi^2   + \sqrt{\frac{d}{n}} \eta,
\label{veq-noise}
\end{equation}
where $\eta$ satisfies $\bra \eta (t) \eta(t') \ket=\delta(t-t')$,
and $d$ is a constant. Here,  we  have ignored effects of 
fluctuations of $J_3$ and the variable dependence of noise 
intensity. It is a non-trivial mathematical problem to clarify whether 
these simplifications are allowed in the 
description of critically divergent fluctuations.
\footnote{Some trajectories satisfy $z(\infty) \ge \zc$,  even when $R < \Rc$.
Such a behavior can be described by (\ref{langevinz}), but 
not by (\ref{veq-noise}). Therefore, the argument below cannot be applied 
to the calculation of $\bra z(\infty) \ket$, for example. Nevertheless, 
we expect that statistical properties of $\theta$ are described 
by (\ref{veq-noise}) 
when we restrict the trajectories $z(\infty) \simeq 0$.} 

Once we are allowed to use  (\ref{veq-noise}), we can determine
the value of the exponent $\nus$ as follows. We set $\epsilon=0$
and write the weight for trajectories 
$[\phi]=(\phi(t))_{0 \le t \le \infty}$:
\begin{equation}
{\cal P}( [\phi] )= \frac{1}{Z}\exp 
\left[   
-\frac{n}{2d} \int dt 
\{ (\partial_t \phi+ b  \phi^2   )^2- \frac{d}{n}  2 b \phi \}
\right],
\label{measure}
\end{equation}
where the last term originates from the Jacobian term
associated with the transformation from $\eta$ to $\phi$.
Now, we define a new scaled variable $\Phi(s)$  
by $\phi(t)= n^{-1/3} \Phi(s)$ with a scaled time $s=n^{-1/3} t$. 
Substituting this into (\ref{measure}), we can 
confirm that the distribution
function of trajectories $\Phi(s)$ is independent of $n$.
This implies that the  time scale near the marginal saddle 
is proportional to $n^{1/3}$. This yields $\zeta'/\nus=1/3$.
Recalling $\zeta'=1/2$, we have arrived at $\nus=3/2$.
From (\ref{rel}), we also obtain $\gamma'=5/2$.
The result is summarized as follows.
\begin{eqnarray}
\bra \theta \ket  &=& n^{1/3} f_1( n^{2/3} \epsilon) , 
\label{sca11} \\
\chi_\theta  &=& n^{5/3} f_2( n^{2/3} \epsilon).
\label{sca22} 
\end{eqnarray}
From these, $\bra \theta \ket \simeq \ep^{-1/2}$ and
$\chi_\theta\simeq \ep^{-5/2}$ in the regime  $O(n^{-2/3}) \ll
\ep  \ll O(n^0) $.
It should be noted that (\ref{sca11}) and (\ref{sca22}) have been 
confirmed numerically for a simple stochastic differential equation 
whose local form near the marginal saddle is equivalent to 
(\ref{veq-noise}) \cite{Iwata3}. 

Now, using this result, we calculate $\bra z(t) \ket$ and $\chi_z(t)$
in the regime $O(n^{-2/3}) \ll
\ep  \ll O(n^0) $, where $\theta$ is expected to 
obey the Gaussian distribution
\begin{eqnarray}
P(\theta)=\frac{1}{Z_{\theta}}
\e^{-n \frac{\lr{\theta-\bra \theta\ket}^2}{2\chi_\theta}}
\label{prob_thetas}
\end{eqnarray}
with the normalization constant $Z_\theta$.
Defining the Fourier transform of $\zu(t)$ as 
\begin{equation}
\zu(t)=\int \frac{d\omega}{2 \pi} \tildezu(\omega)
\e^{i \omega t},
\label{zexp}
\end{equation}
we write approximate expressions
\begin{equation}
\bra z(t) \ket
\simeq  \int \frac{d\omega}{2 \pi}
\tildezu(\omega) \e^{i\omega t} 
\bra \e^{-i \omega \theta }\ket ,
\label{form1:1}
\end{equation}
and 
\begin{equation}
\bra z(t)^2 \ket
\simeq  \int \frac{d\omega}{2 \pi}
\int \frac{d\omega'}{2 \pi}
\tildezu (\omega) \tildezu (\omega')
\e^{i(\omega+\omega')t} 
\bra \e^{-i(\omega+\omega') \theta }\ket .
\label{form2:1}
\end{equation}
The Gaussian distribution (\ref{prob_thetas}) immediately leads us
to 
\begin{equation}
\bra z(t) \ket
\simeq
\int \frac{d\omega}{2 \pi}
\tildezu (\omega)
\e^{i \omega (t-\bra \theta\ket)} 
\e^{ -\frac{\chi_\theta \omega^2}{2n}},
\label{ave_phi}
\end{equation}
and 
\begin{equation}
\bra z(t)^2 \ket
\simeq  \int \frac{d\omega}{2 \pi}
\int \frac{d\omega'}{2 \pi}
\tildezu (\omega)\tildezu (\omega')
\e^{i(\omega+\omega') (t-\bra \theta\ket)}
\e^{ -\frac{\chi_\theta(\omega+\omega')^2}{2n}}.
\label{form2:3}
\end{equation}
Using these, we can derive 
\begin{eqnarray}
\chi_z(t)
&\simeq & 
\chi_\theta \sum_{k=1}^\infty
\left( \frac{1}{k!} 
\left( \sqrt{\frac{\chi_\theta}{n}} \right)^{k-1}
\partial_t^k \bra z(t) \ket \right)^2 .
\label{form2:3-2}
\end{eqnarray}

Here, let $\tauw=\sqrt{\chi_\theta/n}$ be the width 
of the distribution of $\theta$. We expect that 
$\bra z(t) \ket $ can be estimated as $\zu(t-\bra \theta \ket)$
in the regime $\tauw \ll 1$,  where  this regime is 
expressed as $ O(n^{-2/5}) \ll \ep \ll O(n^0)$. 
Then, from (\ref{form2:3-2}), we obtain
\begin{equation}
\chi_z(t) \simeq \chi_\theta  (\partial_t \zu(t) )^2,
\end{equation}
from which we find that  $\chi_z(t)$ takes a maximum at 
$t=t_*$, where $t_* \simeq \bra \theta \ket \simeq \ep^{-1/2}$ 
and $\chi_z(t_*) \simeq \chi_\theta  \simeq \ep^{-5/2}$.
Since the fluctuation intensity is defined as the value of 
$\chi_z(t)$ in  the limit  $n \to \infty$ with small $\epsilon$ fixed,
we conclude that $\zeta=1/2$ and $\gamma=5/2$. 
The behavior of $\chi(\tau)$ in the regime 
$ O(n^{-2/3}) \ll \ep \ll O(n^{-2/5}) $, which is 
described by (\ref{form2:3-2}), seems complicated. 
We conjecture that there is no power-law behavior 
in this regime.
 
%%%%%%%%%%%%%%%%%%%%%%%%%%%%%%%%%%%%%%%%%%%%%%%%%%%%
\section{Concluding remarks}\label{Conclusion}    %
%%%%%%%%%%%%%%%%%%%%%%%%%%%%%%%%%%%%%%%%%%%%%%%%%%%%

We have demonstrated  that the edge deletion processes
of random graphs  exhibit the saddle-node bifurcation in the deterministic
limit, as shown in (\ref{veq}).  The discontinuous transition of 
$\bra \bar h(t=\infty) \ket $, from $0$ to $Q(\zc)$, is understood from 
the nature of the  bifurcation of trajectories of $z(t)$. (See (\ref{35})
for the transformation from $z$ to $\bar h$ in the deterministic 
description.)  We can also understand the  divergent behavior of
$\chi(t)$ on the basis of critical fluctuations of  exit
time from the marginal saddle associated with the saddle-node bifurcation.

The numerical analysis of the power-law divergences is quite difficult, 
although the increasing trends of $\chi(\tau)$ and $\tau$ are easily 
observed,  as already shown in Fig. \ref{fig1}. For example, consider
the power-law  divergence $\chi(\tau)\simeq 
\ep^{-5/2}$ in the regime $0.01 \le \ep \le 0.1$. In this case, 
we need to investigate the system with $n \gg 10^5$. However, since  
our computational algorithm does not involve any tactical 
steps, we cannot perform numerical experiments of
such  a large system.

\begin{figure}[htbp]
\begin{center}
\begin{tabular}{cc}
\includegraphics[width=7cm]{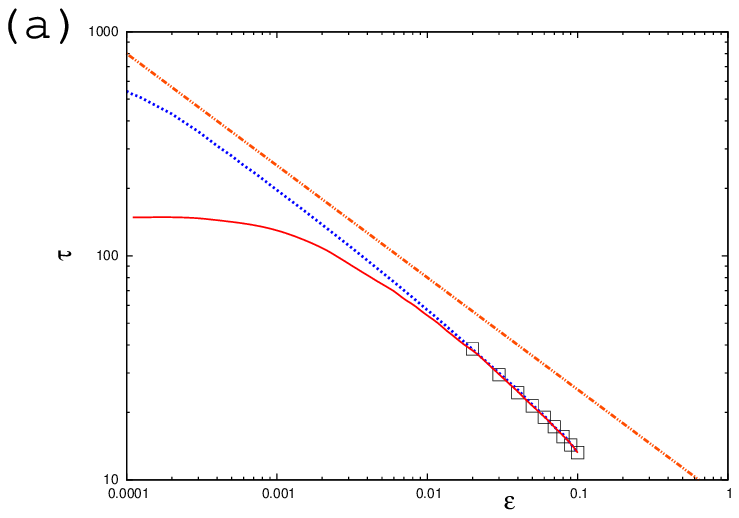}
\includegraphics[width=7cm]{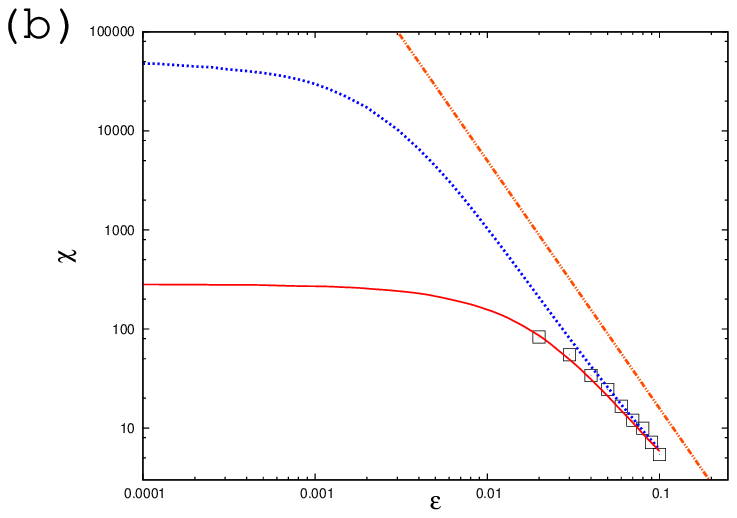}
\end{tabular}
\end{center}
\caption{(a) $ \tau $  as a function of $\epsilon $. 
(b) $ \chi(\tau) $  as a function of $\epsilon$.  
The square symbols represent the numerical results of the 
$k$-core percolation dynamics with $n=2^{13}$. The solid
and dotted curves represent the numerical results of a Langevin 
equation with a small noise intensity that corresponds to 
the cases $n=2^{13}$ and $n\simeq 2^{22}/3$, respectively,
in the $k$-core percolation problem.
The guide lines represent $\tau=8\ep^{-0.5}$ 
and $\chi=0.1 \ep^{-2.5}$.}
\label{fig3}
\end{figure} 

Nevertheless, in Fig. \ref{fig3}, we present the numerical result of 
random graphs on  $n=2^{13}$ vertices. The square symbols represent
$\tau$ and $\chi(\tau)$ for several values of $\epsilon$. 
In order to complement the numerical data, we also display the results 
of numerical simulations of  a simple Langevin  equation 
whose local form near the marginal saddle is equivalent to (\ref{veq-noise}). 
The equation is 
$\partial_t \phi= -\phi((\phi-1)^2+\epsilon)+\sqrt{2T}\xi$,
where $\bra \xi(t)\xi(t')\ket =\delta(t-t')$ and $T$ is 
the noise intensity which is expected to be proportional to 
$1/n$ in the present problem.
The solid curve corresponds to the case $T=2^{-11}/3$,
whose value is chosen such that the square symbols are on the 
solid curves. Then, by decreasing the noise intensity to 
$T=2^{-20}$, which might correspond to the $k$-core problem 
with  $n\simeq 2^{22}/3$, 
we obtain the dotted curve.
As discussed theoretically, the power-law behavior of $\chi \simeq 
\epsilon^{-5/2}$ is observed in the regime $O(n^{-2/5}) \ll \ep \ll O(n^0) $, 
and  $\tau \simeq \epsilon^{-1/2}$ is observed in the regime 
$O(n^{-2/3}) \ll \ep \ll O(n^0) $. 

With regard to finite size effects, we mention that the probability of finding
trajectories satisfying $h(\infty) \not =0$ is given  by a universal 
function of $n^{1/2} (\ep-2 n^{-2/3})$.\footnote{This result was confirmed 
numerically by the direct simulations of the dynamics we consider.
See also Ref. \cite{Dembo} as a mathematical argument.} This implies 
that the system  behavior in the regime $ \ep \le O(n^{-1/2})$ is 
qualitatively different from that in the regime $ \ep \ge O(n^{-1/2})$. 
Theoretically, in order to describe the crossover around $O(n^{-1/2})$, 
we need to analyze (\ref{langevinz}), not (\ref{veq-noise}). 
When we are interested in the relaxation behavior, we should 
focus on the regime $\ep \ge O(n^{-1/2})$. 

It is worthwhile to note that $\chi_\theta$ exhibits the simpler behavior  
than $\chi(\tau)$. We therefore conjecture that $\chi_\theta$ 
is a more fundamental quantity than $\chi(t)$. We also mention 
that the  critical behavior of exit time from a marginal
saddle is observed in a coupled oscillator model \cite{Ohta} 
related to neuronal avalanches \cite{NA}. (See also Ref. \cite{Lindner}.) 
It is an interesting subject to find other examples belonging
to the same universality class.

Although the $k$-core percolation is not directly related to jamming 
transitions, our results might provide a suggestion for future studies 
on jamming transition. As one example of such a study, 
we  may theoretically consider the numerical result obtained for the
jamming transition in the Frederic-Andersen  model in a random graph
\cite{Sellitto}, because  the $k$-core percolation dynamics is
regarded as an irreversible version of a kinetic constraint model.
As another direction of study, one may analyze fluctuations of  
exit time in more general jamming systems. The important example is
the application to the spherical $p$-spin glass model, for which 
the mode coupling theory is believed to be exact \cite{Crisanti}. 
Since the transition described by this theory is interpreted 
as a variant of saddle-node bifurcation \cite{Iwata4}, we
might discuss the divergent behavior of the so-called nonlinear
susceptibility $\chi_4$ on the basis of the exit time from 
the plateau regime.

Finally, we consider the $k$-core percolation in
finite dimensional systems. 
In general, one may conjecture that a transition is smeared in a  
manner similar to bootstrap percolation problems \cite{AL}. (See Refs. 
\cite{Harris} and \cite{parisi-rizzo} for attempts of studying the $k$-core 
percolation in finite dimensional systems.)  From our viewpoint, as the first 
stage of a study on finite dimensional systems,  we  should identify the 
upper-critical dimension $\dc$ for a diffusively coupled  model of a 
simple stochastic system undergoing a saddle-node bifurcation.
With regard to this problem, we point out that $\nus$ might 
be related to the exponent $\nu$ that characterizes the 
divergence of the length scale as $\nu=\nus/\dc$ \cite{Botte}. 
Furthermore, in the next stage of studying finite
dimensional systems, we should consider an equivalence or inequivalence
between such a coupled model and the  $k$-core percolation dynamics in a 
finite dimensional lattice. The analysis developed in the present study
may be useful in this consideration.

%%%%%%%%%%%%%%%%%%%%%%%%%%%%%%%%%%%%%
% acknowledgment                    %
%%%%%%%%%%%%%%%%%%%%%%%%%%%%%%%%%%%%%

\ack

The authors express special thanks to G. Biroli for his suggestion  
that  the dynamics of $k$-core percolation in a random  graph may be 
related to a saddle-node bifurcation.  They also thank G. Biroli (again),
H. Ohta, and H. Tasaki for many useful comments, including the introduction 
of important references. This work was supported by a grant from 
the Ministry of Education, Science, Sports and Culture of Japan, 
No. 19540394. Mami Iwata acknowledges the support  by Hayashi 
memorial foundation for female natural scientists.

\end{document}